\documentclass[onecolumn]{revtex4}
\usepackage{amsfonts}
\usepackage{amsmath}
\usepackage{amssymb}
\usepackage{charter}
\usepackage{graphicx}

\begin{document}

\title{The generalized Hamilton principle and non-Hermitian quantum theory}
\author{Xiang-Yao Wu$^{a,\dag }$, Ben-Shan Wu$^{a}$, Meng Han$^{a}$, Ming-Li
Ren$^{a}$}
\address{Heng-Mei Li$^{b}$, Hong-Chun Yuan$^{c}$, Hong Li$^{d}$ and Si-Qi
Zhang$^{d}$}
\affiliation{$^{a}$Institute of Physics, Jilin Normal University, Siping 136000 China;\\
$^{b}$School of Information Engineering, Changzhou Vocational Institute of
Mechatronic Technology, Changzhou 213164, China\\
$^{c}$School of Electrical and Information Engineering, Changzhou Institute
of Technology, Changzhou 213032, China\\
$^{d}$Institute for Interdisciplinary Quantum Information Technology, Jilin
Engineering Normal University, Changchun 130052, China\\
$^{\dag }$E-mail: wuxy65@126.com}

\begin{abstract}
The Hamilton principle is a variation principle describing the isolated and
conservative systems, its Lagrange function is the difference between
kinetic energy and potential energy. By Feynman path integration, we can
obtain the Hermitian quantum theory, i.e., the standard Schrodinger
equation. In this paper, we have given the generalized Hamilton principle,
which can describe the open system (mass or energy exchange systems) and
nonconservative force systems or dissipative systems. On this basis, we have
given the generalized Lagrange function, it has to do with the kinetic
energy, potential energy and the work of nonconservative forces to do. With
the Feynman path integration, we have given the non-Hermitian quantum theory
of the nonconservative force systems. Otherwise, we have given the
generalized Hamiltonian function for the particle exchanging heat with the
outside world, which is the sum of kinetic energy, potential energy and
thermal energy, and further given the equation of quantum thermodynamics.

\textbf{Keywords:} Hamilton principle; generalized Hamilton principle;
nonconservative systems; non-Hermitian quantum theory.

\textbf{PACS:} 03.65.-w, 05.70.Ce, 05.30.Rt
\end{abstract}

\maketitle

\section{Introduction}

In quantum mechanics, each classical physical quantity corresponds to an
operator, and the operator has a real eigenvalue, which is guaranteed by the
Hermitian operator. The Hermitian operator has always been generally
considered to represent observable measurements. In fact, in quantum
mechanics, it is only necessary to guarantee the observability of the
mechanical quantity, but not to guarantee that its operator must be
Hermitian, that is, observable measurement may also be non-Hermitian. In
1947, in order to solve the divergence problem in the field theory, Pauli
used the indeterminate metric to put forward the theory of the non-Hermitian
operator and its self-consistent inner product, which was derived from a
field quantization method proposed by Dirac \cite{1,2}. In order to maintain
the unitary nature of the $S$ matrix, Lee and Wick applied the non-Hermitian
view to quantum electrodynamics \cite{3}. Later, in different fields,
numerous studies have proved that under certain conditions, the
non-Hermitian Hamiltonian quantum has a real number energy spectrum \cite%
{4,5,6,7}. In 1998, the author Bender proposed the space-time inverse
symmetry (PT symmetry) quantum mechanics, which made the non-Hermitian
quantum mechanics have a great leap forward \cite{8,9}. The non-Hermitian $PT
$ symmetric Hamilton do not violate the physical principles of quantum
mechanics and have real eigenvalues. Over the past decade $PT$ symmetric
quantum theory has been developed into a variety of studies, including field
theory and high-energy particle physics. Recently, preliminary studies on $PT
$ symmetric systems under optical structures have been carried out.

The quantum theory of non-Hermitian is described dissipative systems and
open systems, their unique properties have attracted fast growing interest
in the last two decades \cite{10,11,12,13,14}, especially those empowered by
parity-time symmetry \cite{15}. While the non-Hermitian quantum theories is
still under intense investigation, its application in different fields has
led to a plethora of findings, ranging from nonlinear dynamics\cite{16},
atomic physics \cite{17}, photonics \cite{18}, acoustics \cite{19},
microwave \cite{20}, electronics \cite{21}, to quantum information science%
\cite{22}.

In this paper, we have given the generalized Hamilton principle, which can
describe the open system (mass or energy exchange systems) and
nonconservative force systems or dissipative systems. On this basis, we have
given the generalized Lagrange function, it has to do with the kinetic
energy, potential energy and the work of nonconservative forces to do. With
the Feynman path integration, we have given the non-Hermitian quantum theory
of the nonconservative force systems. Otherwise, we have given the
generalized Hamiltonian function for the particle exchanging heat with the
outside world, which is the sum of kinetic energy, potential energy and
thermal energy, and further given the equation of quantum thermodynamics.

\section{The Hamilton principle for the conservative system}

In a mechanical system, the constraints that limit its position and speed
can be written as equations
\begin{equation}
f(\vec{r}_{i},\dot{\vec{r}}_{i},t)=0,\hspace{0.29in}(i=1,2,\cdots ,h)
\label{1}
\end{equation}%
the number of constraints equations are $h$. For the mechanical system of $N$
free particles, their degree of freedom is $3N$, when they are restricted by
$h$ constraints of equation (\ref{1}), we can select $3N-h$ generalized
coordinates $q_{1},q_{2},\cdots ,q_{3N-h}$, the position vector $\vec{r}_{i}$
can be written as
\begin{equation}
\vec{r}_{i}=\vec{r}_{i}(q_{1},q_{2},\cdots ,q_{3N-h},t),\hspace{0.29in}%
(i=1,2,\cdots ,N)  \label{2}
\end{equation}%
the generalized coordinates $q_{i}$ constitute the configuration space of $%
3N-h$ dimension
\begin{equation}
\vec{q}=[q_{1},q_{2},\cdots ,q_{3N-h}],  \label{3}
\end{equation}%
the virtual displacement are
\begin{equation}
\delta \vec{q}=[\delta q_{1},\delta q_{2},\cdots ,\delta q_{3N-h}],
\label{4}
\end{equation}%
the generalized velocity is
\begin{equation}
\dot{\vec{q}}=\frac{d\vec{q}}{dt}=[\dot{q_{1}},\dot{q_{1}},\cdots ,\dot{q}%
_{3N-h}],  \label{5}
\end{equation}%
where $\dot{q}_{i}=\frac{dq_{i}}{dt}$.

With Eq. (\ref{2}), we have
\begin{equation}
\delta \vec{r}_{i}=\sum_{j}\frac{\partial \vec{r}_{i}}{\partial q_{j}}\delta
q_{j},  \label{6}
\end{equation}
with Eq. (\ref{6}), we can calculate the virtual work of active force $\vec{F%
}_{i}$, it is
\begin{equation}
\delta \omega =\sum_{i}^{N}\vec{F}_{i}\cdot \delta r_{i}=\sum_{i}\vec{F}%
_{i}\cdot \sum_{j}\frac{\partial \vec{r}_{i}}{\partial q_{j}}\delta
q_{j}=\sum_{j}(\sum_{i}\vec{F}_{i}\cdot \frac{\partial \vec{r}_{i}}{\partial
q_{j}})\delta q_{j}=\sum_{j}Q_{j}\delta q_{j},  \label{7}
\end{equation}%
the generalized force $Q_{j}$ is
\begin{equation}
Q_{j}=\sum_{i}\vec{F}_{i}\cdot \frac{\partial \vec{r}_{i}}{\partial q_{j}}.
\label{8}
\end{equation}%
If the generalized force $Q_{j}$ is conservative force, Eq. (7) becomes
\begin{equation}
\delta \omega =\sum_{j}Q_{j}\delta q_{j}=-\delta U,  \label{9}
\end{equation}%
where $U$ is the potential energy.

In rectangular coordinates, there is
\begin{equation}
\vec{F}=-\nabla U,  \label{10}
\end{equation}%
and the component is
\begin{equation}
F_{i}=-\frac{\partial U}{\partial x}.  \label{11}
\end{equation}%
In the following, we should study the system motion from time $t_{1}$ to $%
t_{2}$, the $T$ is the system kinetic energy, there is
\begin{equation}
\int_{t_{1}}^{t_{2}}Tdt=\int_{t_{1}}^{t_{2}}T(q_{i},\dot{q}_{i},t)dt,
\label{12}
\end{equation}%
where $T=\sum_{i}\frac{1}{2}m_{i}\vec{v}_{i}^{2}$.

The variation of Eq.(\ref{12}) is
\begin{equation}
\delta \int_{t_{1}}^{t_{2}}Tdt=\int_{t_{1}}^{t_{2}}\delta
Tdt=\int_{t_{1}}^{t_{2}}\sum_{i}m_{i}{\vec{v_{i}}}\cdot \delta \vec{v}_{i}dt,
\label{13}
\end{equation}%
with $\vec{v}_{i}=\frac{d\vec{r}_{i}}{dt}$, we have
\begin{equation}
\delta \vec{v}_{i}=\frac{d(\delta \vec{r}_{i})}{dt}.  \label{14}
\end{equation}%
Thus, Eq. (\ref{13}) becomes
\begin{equation}
\delta \int_{t_{1}}^{t_{2}}Tdt=\int_{t_{1}}^{t_{2}}\sum_{i}m_{i}\vec{v}%
_{i}\cdot \frac{d(\delta \vec{r}_{i})}{dt}dt=\sum_{i}m_{i}\vec{v}_{i}\cdot
\delta \vec{r}_{i}|_{t_{1}}^{t_{2}}-\int_{t_{1}}^{t_{2}}\sum_{i}m_{i}\dot{%
\vec{v_{i}}}\cdot \delta \vec{r}_{i}dt,  \label{15}
\end{equation}%
i.e.,
\begin{equation}
\delta \int_{t_{1}}^{t_{2}}Tdt+\int_{t_{1}}^{t_{2}}\sum_{i}m_{i}\dot{\vec{%
v_{i}}}\cdot \delta \vec{r}_{i}dt=\sum_{i}m_{i}\dot{\vec{v_{i}}}\cdot \delta
r_{i}|_{t_{1}}^{t_{2}}.  \label{16}
\end{equation}
According to $\vec{F}_{i}=m_{i}\dot{\vec{v_{i}}}$ and $\delta \omega
=\sum_{i}\vec{F}_{i}\cdot \delta r_{i}$, we have
\begin{equation}
\delta \int_{t_{1}}^{t_{2}}Tdt+\int_{t_{1}}^{t_{2}}\sum_{i}\vec{F}_{i}\cdot
\delta \vec{r}_{i}dt=\sum_{i}m_{i}\dot{\vec{v}}_{i}\cdot \delta
r_{i}|_{t_{1}}^{t_{2}}  \label{17}
\end{equation}%
and
\begin{equation}
\delta \int_{t_{1}}^{t_{2}}Tdt+\int_{t_{1}}^{t_{2}}\delta \omega
dt=\sum_{i}m_{i}\dot{\vec{v}}_{i}\cdot \delta r_{i}|_{t_{1}}^{t_{2}}.
\label{18}
\end{equation}%
If the variation of two endpoints are zero, there are
\begin{equation}
\delta q_{j}|_{t_{1}}=\delta q_{j}|_{t_{2}}=0  \label{19}
\end{equation}%
and
\begin{equation}
\delta \vec{r}_{i}|_{t_{1}}=\delta \vec{r}_{i}|_{t_{2}}=0,  \label{20}
\end{equation}%
Eq. (\ref{18}) becomes
\begin{equation}
\delta \int_{t_{1}}^{t_{2}}Tdt+\int_{t_{1}}^{t_{2}}\delta \omega dt=0.
\label{21}
\end{equation}%
As the kinetic energy $T$ is determined by the speed of each moment, there
is
\begin{equation}
\delta \int_{t_{1}}^{t_{2}}Tdt=\int_{t_{1}}^{t_{2}}\delta Tdt.  \label{22}
\end{equation}%
When the active force $F$ is conservative force, the work it does can be
expressed as potential energy $U$, it is
\begin{equation}
\int_{t_{1}}^{t_{2}}\vec{F}\cdot \delta \vec{r}dt=\int_{t_{1}}^{t_{2}}\delta
\omega dt=-\int_{t_{1}}^{t_{2}}\delta Udt=-\delta \int_{t_{1}}^{t_{2}}Udt.
\label{23}
\end{equation}%
Then, Eq. (\ref{21}) becomes
\begin{equation}
\delta \int_{t_{1}}^{t_{2}}(T-U)dt=0,  \label{24}
\end{equation}%
i.e.,
\begin{equation}
\delta \int_{t_{1}}^{t_{2}}Ldt=0,  \label{25}
\end{equation}%
or
\begin{equation}
\delta S=0.  \label{26}
\end{equation}%
Where the Lagrange function $L=T-V$, and the action $S=%
\int_{t_{1}}^{t_{2}}Ldt$. The Eq. (\ref{25}) or (\ref{26}) is the Hamilton
principle for the conservative system.

\section{The generalized Hamilton principle for the nonconservative system}

When the active forces include both conservative force $F_{1}$ and
nonconservative force $F_{2}$, we have
\begin{equation}
\delta \omega =\vec{F}_{1}\cdot \delta \vec{r}+\vec{F}_{2}\cdot \delta \vec{r%
}=\delta \omega _{1}+\delta \omega _{2},  \label{27}
\end{equation}%
and
\begin{equation}
\int_{t_{1}}^{t_{2}}\vec{F}_{1}\cdot \delta \vec{r}dt=\int_{t_{1}}^{t_{2}}%
\delta \omega _{1}dt=-\int_{t_{1}}^{t_{2}}\delta Udt=-\delta
\int_{t_{1}}^{t_{2}}Udt.  \label{28}
\end{equation}%
Substituting Eqs. (\ref{27}) and (\ref{28}) into (\ref{21}), there are
\begin{equation}
\delta \int_{t_{1}}^{t_{2}}(T-U)dt+\int_{t_{1}}^{t_{2}}\delta \omega
_{2}dt=0,  \label{29}
\end{equation}%
and
\begin{equation}
\int_{t_{1}}^{t_{2}}\delta (T-U)dt+\int_{t_{1}}^{t_{2}}\delta \omega
_{2}dt=0,  \label{30}
\end{equation}%
or
\begin{equation}
\int_{t_{1}}^{t_{2}}\delta (T-U+\omega _{2})dt=0.  \label{31}
\end{equation}%
We define generalized Lagrange function $\overline{L}$,
\begin{equation}
\overline{L}=T-U+\omega _{2}=L+\omega _{2},  \label{32}
\end{equation}%
and Eq. (\ref{31}) becomes
\begin{equation}
\int_{t_{1}}^{t_{2}}\delta \overline{L}dt=\int_{t_{1}}^{t_{2}}\delta
(L+\omega _{2})dt=\int_{t_{1}}^{t_{2}}(\delta L+\delta \omega _{2})dt=0.
\label{33}
\end{equation}%
The Eq. (\ref{33}) is called the generalized Hamilton principle for the
nonconservative force system, it is different from the Hamilton principle (%
\ref{25}) for the conservative force system, the Eq. (\ref{33}) contains the
work of nonconservative force, and the variation is inside the integral sign.

From Eq. (\ref{7}), we can give the work of nonconservative forces $%
F_{2i}(i=1,2,\cdots ,N)$, it is
\begin{equation}
\delta \omega _{2}=\sum_{i=1}^{N}\vec{F}_{2i}\cdot \delta \vec{r}%
_{i}=\sum_{j}(\sum_{i}\vec{F}_{2i}\cdot \frac{\partial \vec{r}_{i}}{\partial
q_{j}})\delta q_{j}=\sum_{i}(\sum_{j=1}^{N}\vec{F}_{2j}\cdot \frac{\partial
\vec{r}_{j}}{\partial q_{i}})\delta q_{i}.  \label{34}
\end{equation}%
When there is a single nonconservative force $F_{2}$, there is
\begin{equation}
\delta \omega _{2}=\sum_{i}\vec{F}_{2}\cdot \frac{\partial \vec{r}}{\partial
q_{i}}\delta q_{i}=\sum_{i}F_{2i}\delta q_{i}.  \label{35}
\end{equation}

So, when there are both conservative force $F_{1}$ and nonconservative force
$F_{2}$ for the system, the generalized Lagrange function is
\begin{equation}
\overline{L}=T-U+\omega _{2}=L+\int \vec{F}_{2}\cdot d\vec{r},  \label{36}
\end{equation}%
the generalized action is
\begin{equation}
\overline{S}=\int_{t_{1}}^{t_{2}}\overline{L}dt,  \label{37}
\end{equation}%
and the generalized Hamilton principle is
\begin{equation}
\int_{t_{1}}^{t_{2}}\delta \overline{L}dt=\int_{t_{1}}^{t_{2}}(\delta
L+\delta \omega _{2})dt=0.  \label{38}
\end{equation}%
When there is only nonconservative force $F_{2}$, and there is not
conservative force $F_{1}$ for the system, the generalized Hamilton
principle is
\begin{equation}
\int_{t_{1}}^{t_{2}}\delta \overline{L}dt=\int_{t_{1}}^{t_{2}}(\delta
T+\delta \omega _{2})dt=0,  \label{39}
\end{equation}%
and the generalized Lagrange function is
\begin{equation}
\overline{L}=T+\omega _{2}=T+\int \vec{F}_{2}\cdot d\vec{r}.  \label{40}
\end{equation}

\section{The generalized Hamilton principle for the heat exchange system}

In the mechanical, the change rate of energy is
\begin{equation}
\frac{dE}{dt}=\vec{F}\cdot \vec{v}.  \label{41}
\end{equation}%
For a microcosmic particle, when it exchanges heat $Q$ with the outside
world, there is
\begin{equation}
\frac{dE}{dt}=\frac{dQ}{dt},  \label{42}
\end{equation}%
and the radiant force should be produced, it is
\begin{equation}
\vec{F}\cdot \vec{v}=\frac{dQ}{dt}.  \label{43}
\end{equation}%
When the microcosmic particle absorb heat, $\frac{dQ}{dt}>0$, the radiant
force is $\vec{F}=-k\vec{v}$. When the microcosmic particle deliver heat, $%
\frac{dQ}{dt}<0$, the radiant force is $\vec{F}=k\vec{v}$. The Eq. (\ref{43}%
) should be changed to the following formula
\begin{equation}
\vec{F}\cdot \vec{v}=-\frac{dQ}{dt},  \label{44}
\end{equation}%
i.e.,
\begin{equation}
\vec{F}\cdot d\vec{r}=\vec{F}\cdot \vec{v}\hspace{0.06in}dt=-dQ,  \label{45}
\end{equation}%
then
\begin{equation}
\int \vec{F}\cdot d\vec{r}=-\int dQ=-Q,  \label{46}
\end{equation}%
the radiant force is a nonconservative force. When a microcosmic particle
exchanges heat with the outside world, its generalized Lagrange function is
\begin{equation}
\overline{L}=T-U+\int \vec{F}\cdot d\vec{r}=L-Q,  \label{47}
\end{equation}%
the generalized Hamiltonian function for the heat exchange system is
\begin{equation}
\overline{H}=p\dot{q}-\overline{L}=T+U+Q,  \label{48}
\end{equation}%
and the generalized Hamilton principle for the heat exchange system is
\begin{equation}
\int_{t_{1}}^{t_{2}}\delta \overline{L}dt=\int_{t_{1}}^{t_{2}}(\delta
L-\delta Q)dt=0.  \label{49}
\end{equation}

\section{The generalized Lagrange equation and generalized Hamilton function
for the nonconservative system}

(1) The generalized Lagrange equation for the nonconservative system

For the nonconservative system, the generalized Lagrange function is
\begin{equation}
\overline{L}=T-U+\omega _{2}=L+\omega _{2}=L+\int \vec{F}_{2}\cdot d\vec{r},
\label{50}
\end{equation}%
i.e.,
\begin{equation}
\overline{L}=L(q_{i},\dot{q}_{i},t)+\omega _{2}(\vec{r}(q_{i}),t),
\label{51}
\end{equation}%
the variation of $\overline{L}$ is
\begin{eqnarray}
\delta \overline{L} &=&\delta L(q_{i},\dot{q}_{i},t)+\delta \omega _{2}(\vec{%
r}(q_{i}),t)  \notag \\
&=&\frac{\partial L}{\partial q_{i}}\delta q_{i}+\frac{\partial L}{\partial
\dot{q}_{i}}\delta \dot{q}_{i}+\vec{F}_{2}\cdot \delta \vec{r}  \notag \\
&=&\frac{\partial L}{\partial q_{i}}\delta q_{i}+\frac{\partial L}{\partial
\dot{q}_{i}}\delta \dot{q}_{i}+\vec{F}_{2}\cdot \frac{\partial \vec{r}}{%
\partial {q}_{i}}\delta {q}_{i}  \notag \\
&=&\frac{\partial L}{\partial q_{i}}\delta q_{i}+\frac{\partial L}{\partial
\dot{q}_{i}}\delta \dot{q}_{i}+F_{2i}\cdot \delta {q}_{i},  \label{52}
\end{eqnarray}%
where $\delta \omega _{2}=\vec{F}_{2}\cdot \delta \vec{r}$, $\delta \vec{r}=%
\frac{\partial \vec{r}}{\partial {q}_{i}}\delta {q}_{i}$ and $\vec{F}%
_{2}\cdot \frac{\partial \vec{r}}{\partial {q}_{i}}=F_{2i}$.\newline

Substituting Eq. (\ref{52}) into the generalized Hamilton principle (\ref{38}%
), there is
\begin{equation}
\int_{t_{1}}^{t_{2}}\delta \overline{L}dt=\int_{t_{1}}^{t_{2}}(\frac{%
\partial L}{\partial q_{i}}\delta q_{i}+\frac{\partial L}{\partial \dot{q}%
_{i}}\delta \dot{q}_{i}+F_{2i}\cdot \delta {q}_{i})dt=0.  \label{53}
\end{equation}%
Obviously, there is
\begin{equation}
\int_{t_{1}}^{t_{2}}\frac{\partial L}{\partial \dot{q}_{i}}\delta \dot{q}%
_{i}dt=-\int_{t_{1}}^{t_{2}}\frac{d}{dt}\frac{\partial L}{\partial \dot{q}%
_{i}}\delta q_{i}dt.  \label{54}
\end{equation}%
Substituting Eq. (\ref{54}) into (\ref{53}), we have
\begin{equation}
\frac{d}{dt}\frac{\partial L}{\partial \dot{q}_{i}}-\frac{\partial L}{%
\partial q_{i}}=F_{2i}.  \label{55}
\end{equation}%
The Eq. (\ref{55}) is the generalized Lagrange equation for the
nonconservative system.\newline

(2) The generalized Hamilton function for the nonconservative system\newline

When $L$ and $w_{2}$ do not include time, the time derivative of $\overline{L%
}$ is
\begin{eqnarray}
\frac{d\overline{L}}{dt} &=&\frac{\partial L}{\partial q_{i}}\dot{q}_{i}+%
\frac{\partial L}{\partial \dot{q}_{i}}{\ddot{q}}_{i}+\frac{\partial w_{2}}{%
\partial \vec{r}}\cdot \frac{\partial \vec{r}}{\partial {q}_{i}}\dot{q}_{i}
\notag \\
&=&\frac{\partial L}{\partial q_{i}}\dot{q}_{i}+\frac{\partial L}{\partial
\dot{q}_{i}}{\ddot{q}}_{i}+\vec{F}_{2}\cdot \frac{\partial \vec{r}}{\partial
{q}_{i}}\dot{q}_{i}  \notag \\
&=&(\frac{\partial L}{\partial q_{i}}+F_{2i})\dot{q}_{i}+\frac{\partial L}{%
\partial \dot{q}_{i}}{\ddot{q}}_{i}  \notag \\
&=&\frac{d}{dt}\frac{\partial L}{\partial \dot{q}_{i}}\dot{q}_{i}+\frac{%
\partial L}{\partial \dot{q}_{i}}\ddot{q}_{i}  \notag \\
&=&\frac{d}{dt}(\frac{\partial L}{\partial \dot{q}_{i}}\dot{q}_{i}),
\label{56}
\end{eqnarray}%
where $w_{2}=\int \vec{F}_{2}\cdot d\vec{r}$ and $\vec{F}_{2}=\frac{\partial
w_{2}}{\partial \vec{r}}$.\newline

Using Eq. (\ref{56}), we have
\begin{equation}
\frac{d}{dt}(\frac{\partial L}{\partial \dot{q}_{i}}\dot{q}_{i}-\overline{L}%
)=0,  \label{57}
\end{equation}%
or
\begin{equation}
\frac{\partial L}{\partial \dot{q}_{i}}\dot{q}_{i}-L-w_{2}=\overline{H}%
=constant.  \label{58}
\end{equation}%
As
\begin{equation}
\frac{\partial L}{\partial \dot{q}_{i}}\dot{q}_{i}-L=T+U=H,  \label{59}
\end{equation}%
then
\begin{equation}
\overline{H}=T+U-w_{2}=H-w_{2}.  \label{60}
\end{equation}%
The $\overline{H}$ is called the integral of generalized energy, or
generalized Hamilton function for the nonconservative force system.\newline

(3) The invariance of $\overline{L}$ and the conserved quantity\newline

With Eqs. (\ref{52}) and (\ref{55}), we have
\begin{eqnarray}
\delta \overline{L} &=&\frac{\partial L}{\partial q_{i}}\delta q_{i}+\frac{%
\partial L}{\partial \dot{q}_{i}}\delta \dot{q}_{i}+F_{2i}\cdot \delta {q}%
_{i}  \notag \\
&=&(\frac{\partial L}{\partial q_{i}}+F_{2i})\delta q_{i}+\frac{\partial L}{%
\partial \dot{q}_{i}}\delta \dot{q}_{i}  \notag \\
&=&\frac{d}{dt}\frac{\partial L}{\partial \dot{q}_{i}}\delta \dot{q}_{i}+%
\frac{\partial L}{\partial \dot{q}_{i}}\delta \dot{q}_{i}  \notag \\
&=&\frac{d}{dt}(\frac{\partial L}{\partial \dot{q}_{i}}\delta {q}_{i})=0.
\label{61}
\end{eqnarray}%
By the invariance of $\overline{L}$ ($\delta \overline{L}=0$), we can obtain
the conserved quantity for the nonconservative system%
\begin{equation}
\frac{\partial L}{\partial \dot{q}_{i}}\delta {q}_{i}=constant,  \label{62-1}
\end{equation}
which is the same as the conservative system.

\section{The generalized Lagrange equation and generalized Hamilton function
for the heat exchange system}

(1) The generalized Lagrange equation for the heat exchange system\newline

In Eq. (\ref{47}), the generalized Lagrange function for the heat exchange
system is
\begin{equation}
\overline{L}=T-U-Q=L-Q,  \label{63}
\end{equation}%
In section 8 (Eq. (\ref{91})), we have given the microcosmic heat $Q=TS$,
then the Eq. (\ref{63}) becomes
\begin{equation}
\overline{L}=L-TS,  \label{64}
\end{equation}%
i.e.,
\begin{equation}
\overline{L}=L({q}_{i},\dot{q}_{i},t)-ST({q}_{i},\dot{q}_{i},t).  \label{65}
\end{equation}%
When $L$ and $T$ do not include time, the variation of $\overline{L}$ is
\begin{equation}
\delta \overline{L}=\frac{\partial L}{\partial q_{i}}\delta q_{i}+\frac{%
\partial L}{\partial \dot{q}_{i}}\delta \dot{q}_{i}-S\frac{\partial T}{%
\partial q_{i}}\delta q_{i}.  \label{66}
\end{equation}%
Substituting Eq. (\ref{66}) into the generalized Hamilton principle (\ref{38}%
), there is
\begin{eqnarray}
\int_{t_{1}}^{t_{2}}\delta \overline{L}dt &=&\int_{t_{1}}^{t_{2}}(\frac{%
\partial L}{\partial q_{i}}\delta q_{i}+\frac{\partial L}{\partial \dot{q}%
_{i}}\delta \dot{q}_{i}-S\frac{\partial T}{\partial q_{i}}\delta q_{i})dt
\notag \\
&=&\int_{t_{1}}^{t_{2}}(\frac{\partial L}{\partial q_{i}}-\frac{d}{dt}\frac{%
\partial L}{\partial \dot{q}_{i}}-S\frac{\partial T}{\partial q_{i}})\delta
q_{i}dt=0.  \label{67}
\end{eqnarray}%
When the $\delta q_{i}$ is arbitrary, we obtain
\begin{equation}
\frac{\partial L}{\partial q_{i}}-\frac{d}{dt}\frac{\partial L}{\partial
\dot{q}_{i}}-S\frac{\partial T}{\partial q_{i}}=0.  \label{68}
\end{equation}%
The Eq. (\ref{68}) is the generalized Lagrange equation for the heat
exchange system.\newline

(2) The generalized Hamilton function for the heat exchange system\newline

When $L$ and $T$ do not include time, the time derivative of $\overline{L}$
is
\begin{eqnarray}
\frac{d\overline{L}}{dt} &=&\frac{\partial L}{\partial q_{i}}\dot{q}_{i}+%
\frac{\partial L}{\partial \dot{q}_{i}}{\ddot{q}}_{i}-S\frac{\partial T}{%
\partial q_{i}}\dot{q}_{i}  \notag \\
&=&(\frac{\partial L}{\partial q_{i}}-S\frac{\partial T}{\partial q_{i}})%
\dot{q}_{i}+\frac{\partial L}{\partial \dot{q}_{i}}{\ddot{q}}_{i}  \notag \\
&=&\frac{d}{dt}\frac{\partial L}{\partial \dot{q}_{i}}\dot{q}_{i}+\frac{%
\partial L}{\partial \dot{q}_{i}}\ddot{q}_{i}  \notag \\
&=&\frac{d}{dt}(\frac{\partial L}{\partial \dot{q}_{i}}\dot{q}_{i}),
\label{69}
\end{eqnarray}%
With Eq. (\ref{69}), we have
\begin{equation}
\frac{d}{dt}(\frac{\partial L}{\partial \dot{q}_{i}}\dot{q}_{i}-\overline{L}%
)=0,  \label{70}
\end{equation}%
or
\begin{equation}
\frac{\partial L}{\partial \dot{q}_{i}}\dot{q}_{i}-L+TS=\overline{H}%
=constant.  \label{71}
\end{equation}%
For
\begin{equation}
\frac{\partial L}{\partial \dot{q}_{i}}\dot{q}_{i}-L=T+U=H,  \label{72}
\end{equation}%
then
\begin{equation}
\overline{H}=T+U+Q=H+Q=H+TS.  \label{73}
\end{equation}%
The $\overline{H}$ is called the integral of generalized energy, or
generalized Hamilton function for the heat exchange system.\newline

(3) The invariance of $\overline{L}$ and the conserved quantity\newline

In Eqs. (\ref{66}) and (\ref{68}), we have
\begin{eqnarray}
\delta \overline{L} &=&\frac{\partial L}{\partial q_{i}}\delta q_{i}+\frac{%
\partial L}{\partial \dot{q}_{i}}\delta \dot{q}_{i}-S\frac{\partial T}{%
\partial q_{i}}\delta q_{i}  \notag \\
&=&(\frac{\partial L}{\partial q_{i}}-S\frac{\partial T}{\partial q_{i}}%
)\delta q_{i}+\frac{\partial L}{\partial \dot{q}_{i}}\delta \dot{q}_{i}
\notag \\
&=&\frac{d}{dt}\frac{\partial L}{\partial \dot{q}_{i}}\delta q_{i}+\frac{%
\partial L}{\partial \dot{q}_{i}}\delta \dot{q}_{i}  \notag \\
&=&\frac{d}{dt}(\frac{\partial L}{\partial \dot{q}_{i}}\delta {q}_{i})=0.
\label{74}
\end{eqnarray}%
By the invariance of $\overline{L}$ ($\delta \overline{L}=0$), we can obtain
the conserved quantity for the heat exchange system
\begin{equation}
\frac{\partial L}{\partial \dot{q}_{i}}\delta {q}_{i}=constant,  \label{75}
\end{equation}%
which is the same as the conservative system.\newline

In the above, we have given the generalized Hamilton principle for the
nonconservative force and the heat exchange system. On this basis, we
further given the generalized Lagrange function and generalized Hamilton
function for the nonconservative force and the heat exchange system. With
the results, we shall study the non-Hermitian quantum theory for the
nonconservative force and the heat exchange microcosmic system.

\section{The non-Hermitian quantum theory for the nonconservative force
system}

With the generalized Hamilton principle and generalized Lagrange function,
we will deduce the non-Hermitian quantum theory for the nonconservative
force system by the approach of path integral, and the path integral formula
is
\begin{equation}
\Psi (\vec{r},t^{\prime })=\int \exp [\frac{i}{\hbar }\int_{t}^{t^{\prime }}%
\overline{L}(\dot{\vec{r}}(\tau ),{\vec{r}}(\tau ),\tau )d\tau ]D[\vec{r}%
(t)]\Psi (\overrightarrow{r}^{\prime },t)d\overrightarrow{r}^{\prime }.
\label{76}
\end{equation}%
In Eq. (\ref{76}), the generalized Lagrange function $\overline{L}$ is
\begin{equation}
\overline{L}=T-U+\omega _{2}=L+\omega _{2}=L+\int \vec{F}\cdot d\vec{r},
\label{77}
\end{equation}%
where the force $\vec{F}$ is the nonconservative force. The Eq. (\ref{76})
gives the wave function at a time $t^{\prime }$ in terms of the wave
function at a time $t$. In order to obtain the differential equation, we
apply this relationship in the special case that the time $t^{\prime }$
differs only by an infinitesimal interval $\varepsilon $ from $t$. For a
short interval $\varepsilon $ the action is approximately $\varepsilon $
times the Lagrangian for this interval, we have
\begin{equation}
\Psi (\vec{r},t+\varepsilon )=\int \frac{d\overrightarrow{r}^{\prime }}{A^{3}%
}\exp [\frac{i\varepsilon }{\hbar }\overline{L}(\frac{\vec{r}-%
\overrightarrow{r}^{\prime }}{\varepsilon },\frac{\vec{r}+\overrightarrow{r}%
^{\prime }}{2},\frac{t^{\prime }+t}{2})]\Psi (\overrightarrow{r}^{\prime
},t),  \label{78}
\end{equation}%
where $A$ is a normalization constant.

Substituting Eq. (\ref{77}) into (\ref{78}), there is
\begin{equation}
\Psi (\vec{r},t+\varepsilon )=\int \frac{d\overrightarrow{r}^{\prime }}{A^{3}%
}\exp [\frac{i\varepsilon }{\hbar }(\frac{m}{2}(\frac{\vec{r}-%
\overrightarrow{r}^{\prime }}{\varepsilon })^{2}-V(\frac{\vec{r}+%
\overrightarrow{r}^{\prime }}{2},\frac{t^{\prime }+t}{2})+\int_{%
\overrightarrow{r}^{\prime }}^{\vec{r}}\vec{F}\cdot d\overrightarrow{r}%
^{\prime \prime })]\Psi (\overrightarrow{r}^{\prime },t).  \label{79}
\end{equation}%
In macroscopic field, the frictional force and adhere force are
non-conservative force, and the non-conservative force $\vec{F}$ is directly
proportional to velocity $\vec{v}$, their directions are opposite, i.e. $%
\vec{F}=-k\vec{v}$. In microcosmic field, atomic and molecular can also
suffer the action of non-conservative force. In the experiment of
Bose-Einstein condensates, the atomic $Rb^{87}$, $Na^{23}$ and $Li^{7}$ can
be cooled in laser field, since they get the non-conservative force from the
photons.

Substituting $\vec{F}=-k\vec{v}$ into Eq. (\ref{79}), we get
\begin{eqnarray}
\Psi (\vec{r},t+\varepsilon ) &=&\int \frac{d\overrightarrow{r}^{\prime }}{%
A^{3}}\exp [\frac{i\varepsilon }{\hbar }(\frac{m}{2}(\frac{\vec{r}-%
\overrightarrow{r}^{\prime }}{\varepsilon })^{2}-V(\frac{\vec{r}+%
\overrightarrow{r}^{\prime }}{2},\frac{t^{\prime }+t}{2})  \notag \\
&&-k\int_{\overrightarrow{r}^{\prime }}^{\vec{r}}(\frac{\vec{r}-%
\overrightarrow{r}^{\prime }}{\varepsilon })\cdot d\overrightarrow{r}%
^{^{\prime \prime }})]\Psi (\overrightarrow{r}^{\prime },t).  \label{80}
\end{eqnarray}%
The quantity $(\frac{\vec{r}-\overrightarrow{r}^{\prime }}{\varepsilon })^{2}
$ appears in the exponent of the first factor. It is clear that if $%
\overrightarrow{r}^{\prime }$ is appreciably different from $\vec{r}$, this
quantity is very large and the exponential consequently oscillates very
rapidly as $\overrightarrow{r}^{\prime }$ varies. When this factor
oscillates rapidly, the integral over $\overrightarrow{r}^{\prime }$ gives a
very small value. Only if $\overrightarrow{r}^{\prime }$ is near $\vec{r}$
do we get important contributions. For this reason, we make the substitution
$\overrightarrow{r}^{\prime }=\vec{r}+\vec{\eta}$ with the expectation that
appreciable contribution to the integral will occur only for small $\vec{\eta%
}$, and we obtain
\begin{equation}
\Psi (\vec{r},t+\varepsilon )=\int \frac{d\vec{\eta}}{A^{3}}\exp [\frac{%
i\varepsilon }{\hbar }(\frac{m}{2}(\frac{\vec{\eta}}{\varepsilon })^{2}-V(%
\vec{r}+\frac{\vec{\eta}}{2},t+\frac{\varepsilon }{2})-k\int_{%
\overrightarrow{r}^{\prime }}^{\vec{r}}\frac{-\vec{\eta}}{\varepsilon }\cdot
d\overrightarrow{r}^{\prime \prime })]\Psi (\vec{r}+\vec{\eta},t).
\label{81}
\end{equation}%
Now we have
\begin{equation}
\int_{\overrightarrow{r}^{\prime }}^{\vec{r}}\vec{\eta}\cdot d%
\overrightarrow{r}_{\overrightarrow{r}^{\prime }}^{\prime \prime \vec{r}}|%
\vec{\eta}||d\overrightarrow{r}^{\prime \prime }|\cos \theta =|\vec{\eta}%
|\int_{\overrightarrow{r}^{\prime }}^{\vec{r}}|d\overrightarrow{r}^{\prime
\prime }|\cos \theta =|\vec{\eta}|^{2},  \label{82}
\end{equation}%
so that
\begin{equation}
k\int_{\overrightarrow{r}^{\prime }}^{\vec{r}}\frac{-\vec{\eta}}{\varepsilon
}\cdot d\overrightarrow{r}^{\prime \prime }=-\frac{k}{\varepsilon }|\vec{\eta%
}|^{2}=-\frac{k}{\varepsilon }\vec{\eta}^{2}.  \label{83}
\end{equation}%
Substituting Eq. (\ref{83}) into (\ref{81}), we have
\begin{eqnarray}
\Psi (\vec{r},t+\varepsilon ) &=&\int \frac{d\vec{\eta}}{A^{3}}\exp [\frac{%
i\varepsilon }{\hbar }(\frac{m}{2}\frac{\vec{\eta}^{2}}{\varepsilon ^{2}}-V(%
\vec{r}+\frac{\vec{\eta}}{2},t+\frac{\varepsilon }{2})+\frac{k}{\varepsilon }%
\vec{\eta}^{2})]\Psi (\vec{r}+\vec{\eta},t)  \notag \\
&=&\int \frac{d\vec{\eta}}{A^{3}}e^{\frac{im\vec{\eta}^{2}}{2\hbar
\varepsilon }}e^{-\frac{i\varepsilon }{\hbar }V(\vec{r}+\frac{\vec{\eta}}{2}%
,t+\frac{\varepsilon }{2})}e^{+\frac{i}{\hbar }k\vec{\eta}^{2}}\Psi (\vec{r}+%
\vec{\eta},t).  \label{84}
\end{eqnarray}%
After more complex calculation, it is obtained that
\begin{equation}
i\hbar \frac{\partial \Psi (\vec{r},t)}{\partial t}=(-\frac{\hbar ^{2}}{2m}%
\nabla ^{2}+V-i\hbar \frac{3k}{m})\Psi (\vec{r},t)=\hat{H}\Psi (\vec{r},t),
\label{85}
\end{equation}%
where the Hamiltonian $H$ is
\begin{equation}
\hat{H}=-\frac{\hbar ^{2}}{2m}\nabla ^{2}+V-i\hbar \frac{3k}{m}.  \label{86}
\end{equation}%
Obviously, the Hamiltonian $H$ is non Hermitian. The Detailed derivation can
see the Ref.\cite{23}.

\section{The non-Hermitian quantum theory for the thermodynamics}

In classical mechanics, the energy of a macroscopic object is
\begin{equation}
E=\frac{{p}^{2}}{2m}+{V}(r).  \label{87}
\end{equation}%
For a microcosmic particle, when it exchanges heat $Q$ with the outside
world. Using Eq. (\ref{48}) or (\ref{73}), the particle total energy should
be the sum of kinetic energy, potential energy and thermal energy, namely,
\begin{equation}
E=\frac{p^{2}}{2m}+V(r)+Q.  \label{88}
\end{equation}%
In thermodynamics, for the infinitely small processes, the entropy is
defined as
\begin{equation}
dS=\frac{dQ}{T}.  \label{89}
\end{equation}%
For the finite processes, it is
\begin{equation}
Q-Q_{0}=TS-TS_{0}.  \label{90}
\end{equation}%
At temperature $T$, when a particle has the microcosmic entropy $S$, it
should has the thermal potential energy $Q$,
\begin{equation}
Q=TS,  \label{91}
\end{equation}%
and the Eq. (\ref{88}) becomes
\begin{equation}
E=\frac{p^{2}}{2m}+V(r)+TS.  \label{92}
\end{equation}%
Eq. (\ref{92}) is the classical total energy of a microcosmic particle. In
quantum theory, it should become operator form
\begin{equation}
\hat{H}=\frac{\hat{p}^{2}}{2m}+\hat{V}(r)+T\hat{S},  \label{93}
\end{equation}%
where $\hat{H}=i\hbar \frac{\partial }{\partial t}$, $\hat{p}^{2}=-\hbar
^{2}\nabla ^{2}$ and $\hat{S}$ is the microcosmic entropy operator.\newline

At the $i-th$ microcosmic state, the classical microcosmic entropy $S_{Fi}$
and $S_{Bi}$ for Fermion and Bose systems are
\begin{equation}
S_{Fi}=-k_{B}[n_{i}\ln n_{i}+(1-n_{i})\ln (1-n_{i})],  \label{94}
\end{equation}%
and
\begin{equation}
S_{Bi}=-k_{B}[n_{i}\ln n_{i}-(1+n_{i})\ln (1+n_{i})],  \label{95}
\end{equation}%
where $k_{B}$ is the Boltzmann constant, $n_{i}$ is the average particle
numbers of particle in the $i-th$ state. For the Fermion (Bose), $n_{i}\leq 1
$ ($n_{i}\geq 1$).

In quantum theory, the classical microcosmic entropy should become operator.
The microcosmic entropy operator depends on temperature, but it has no the
dimension of temperature, and it is non-Hermitian operator because it has to
do with heat exchange. Therefore, the microcosmic entropy operator includes
the temperature operator $T\frac{\partial }{\partial T}$. Moreover, it has
to do with the state distribution. For the Fermion and Bose systems, the
microcosmic entropy operator $\hat{S}_{Fi}$ and $\hat{S}_{Bi}$ of a particle
in the $i-th$ state can be written as
\begin{equation}
\hat{S}_{Fi}=-k_{B}[n_{i}\ln n_{i}+(1-n_{i})\ln (1-n_{i})]T\frac{\partial }{%
\partial T}=S_{Fi}T\frac{\partial }{\partial T},  \label{96}
\end{equation}%
and
\begin{equation}
\hat{S}_{Bi}=-k_{B}[n_{i}\ln n_{i}-(1+n_{i})\ln (1+n_{i})]T\frac{\partial }{%
\partial T}=S_{Bi}T\frac{\partial }{\partial T}.  \label{97}
\end{equation}%
In addition, we can prove the following operator relation
\begin{equation}
\hat{T}^{+}=\hat{T}=T,  \label{98}
\end{equation}%
\begin{equation}
(-i{\frac{\partial }{\partial T}})^{+}=-i{\frac{\partial }{\partial T}}
\label{99}
\end{equation}%
and%
\begin{equation}
\lbrack \hat{T},\frac{\partial }{\partial T}]=-1.  \label{100}
\end{equation}%
According to Eqs. (\ref{98})-(\ref{100}), we find that the operator $T\frac{%
\partial }{\partial T}$ is non-Hermitian. Thus, the microcosmic entropy
operators (\ref{96}) and (\ref{97}) are also non-Hermitian, and the total
Hamilton operator (\ref{93}) is non-Hermitian and space-time inversion ($PT$%
) symmetry
\begin{equation}
\hat{H}^{+}\neq H,\hspace{0.2in}(PT)H(PT)^{-1}=H.  \label{101}
\end{equation}%
This is because the particle (atom or molecule) exchanges energy with the
external environment, which is an open system, its Hamiltonian operator
should be non-Hermitian.

\section{The Schroding equation with temperature}

With the canonical quantization, $E=i\hbar \frac{\partial }{\partial t}$, $%
\vec{p}=-i\hbar \nabla $, substituting Eq. (\ref{96}) into (\ref{93}), we
can obtain the Schroding equation with temperature
\begin{equation}
i\hbar \frac{\partial }{\partial t}\psi (\vec{r},t,T)=(-\frac{\hbar ^{2}}{2m}%
\nabla ^{2}+V(r)+\sum_{i}S_{Fi}T^{2}\frac{\partial }{\partial T})\psi (\vec{r%
},t,T).  \label{102}
\end{equation}

By separating variables
\begin{equation}
\psi (\vec{r},t.T)=\Psi (\vec{r},T)f(t),  \label{103}
\end{equation}%
we obtain
\begin{equation}
i\hbar \frac{df(t)}{dt}=E_{n}f(t),  \label{104}
\end{equation}%
\begin{equation}
(-\frac{\hbar ^{2}}{2m}\nabla ^{2}+V(r)+f_{n}S_{Fi}T^{2}\frac{\partial }{%
\partial T})\psi _{n}(\vec{r},T)=E_{n}\psi _{n}(\vec{r},T).  \label{105}
\end{equation}%
By separating variables $\Psi _{n}(\vec{r},T)=\Psi _{n}(\vec{r})\phi (T)$,
Eq. (\ref{105}) can be written as
\begin{equation}
-\frac{\hbar ^{2}}{2m}\nabla ^{2}\Psi _{n}(\vec{r})+V(r)\Psi _{n}(\vec{r}%
)=E_{1n}\Psi _{n}(\vec{r}),  \label{106}
\end{equation}%
and%
\begin{equation}
{f_{n}}S_{Fi}T^{2}\frac{\partial }{\partial T}\phi (T)=E_{2n}\phi (T),
\label{107-1}
\end{equation}
where $E_{n}=E_{1n}+E_{2n}$ with $E_{1n}$ being the eigenenergy obtained by
the Schroding equation (\ref{106}) and $E_{2n}$ being the eigenenergy
obtained by the temperature equation (\ref{107-1}), the $n$ expresses the $%
n-th$ energy level, $n_{i}$ is the average particle numbers of the $i-th$
state in the $n-th$ energy level, and $f_{n}$ is the degeneracy of the $n-th$
energy level.

From Eq. (\ref{107-1}), we can obtain that its solution is
\begin{equation}
E_{2n}=f_{n}S_{Fi}T_{0}=-k_{B}f_{n}[n_{i}\ln n_{i}+(1-n_{i})\ln
(1-n_{i})]T_{0},  \label{108}
\end{equation}%
and
\begin{equation}
T^{2}\frac{\partial }{\partial T}\phi (T)=\phi (T)T_{0},  \label{109}
\end{equation}%
where the temperature wave function $\phi (T)$ is
\begin{equation}
\phi (T)=Ae^{-\frac{T_{0}}{T}}  \label{110}
\end{equation}%
with $A$ being the normalization constant and $T_{0}$ being the temperature
constant. The general solution of Eq. (\ref{102}) is
\begin{equation}
\psi (\vec{r},t,T)=\sum_{n}C_{n}\Psi _{n}(\vec{r})\phi _{n}(T)e^{-\frac{i}{%
\hbar }E_{n}t}.  \label{111}
\end{equation}%
For a free particle, its momentum is $\vec{p}$ in the environment of
temperature $T$. Because this is in the determinate state, i.e., the average
particle numbers $n_{i}=\delta _{ij}$, the free particle plane wave solution
and total energy are
\begin{equation}
\psi (\vec{r},t,T)=Ae^{\frac{i}{\hbar }(\vec{p}\cdot \vec{r}-Et+i\frac{T_{0}%
}{T}\hbar )}  \label{112}
\end{equation}%
and
\begin{equation}
E=\frac{p^{2}}{2m}.  \label{113}
\end{equation}

By the accurate measurement the hydrogen atom spectrum, we can determine the
temperature constant $T_{0}$. The hydrogen atom has only one electron
outside the nucleus, the degeneracy of the $n-th$ energy level is $%
f_{n}=n^{2}$. When the electron jumps from $m-th$ energy level to the $n-th$
energy level $(m>n)$ , the transition frequency without temperature
correction (the theoretical calculation with Schroding equation) is
\begin{equation}
\nu _{mn}^{th}=\frac{E_{m}-E_{n}}{h},  \label{114}
\end{equation}%
and the transition frequency with temperature correction is
\begin{equation}
\nu _{mn}(T)=\nu _{mn}^{exp}=\frac{E_{m}(T)-E_{n}(T)}{h},  \label{115}
\end{equation}%
where the energy levels $E_{m}(T)$ and $E_{n}(T)$ are
\begin{equation}
E_{m}(T)=E_{m}-k_{B}f_{m}[m_{i}\ln m_{i}+(1-m_{i})\ln (1-m_{i})]T_{0},
\label{116}
\end{equation}%
and
\begin{equation}
E_{n}(T)=E_{n}-k_{B}f_{n}[n_{i}\ln n_{i}+(1-n_{i})\ln (1-n_{i})]T_{0},
\label{117}
\end{equation}%
respectively. The average particle numbers of every state in the $m-th$ and $%
n-th$ energy levels are $m_{i}=\frac{1}{m^{2}}$ and $n_{i}=\frac{1}{n^{2}}$,
respectively.

Considering Eqs. (\ref{114}) and (\ref{115}), we obtain the temperature
constant $T_{0}$,
\begin{equation}
T_{o}=\frac{h(\nu _{mn}^{exp}-\nu _{mn}^{th})}{k_{B}[\ln \frac{m^{2}}{n^{2}}%
-(m^{2}-1)\ln (1-\frac{1}{m^{2}})+(n^{2}-1)\ln (1-\frac{1}{n^{2}})]},
\label{118}
\end{equation}%
where $h$ is the Planck constant. By measurement transition frequency $\nu
_{mn}^{exp}$, we can determine the temperature constant $T_{0}$. When the
electron jumps from the first excited state $(m=2)$ to ground state $(n=1)$,
the $T_{o}$ is
\begin{equation}
T_{o}=\frac{h(\nu _{21}^{exp}-\nu _{21}^{th})}{k_{B}[4\ln 4-3\ln 3]}.
\label{119}
\end{equation}%
The theory should be tested by the experiments.

\section{Conclusions}

The Hamilton principle is a variation principle describing the isolated and
conservative systems, its Lagrange function is the difference between
kinetic energy and potential energy. By Feynman path integration, we can
obtain the Hermitian quantum theory, i.e., the standard Schrodinger
equation. In this paper, we have given the generalized Hamilton principle,
which can describe the open system (mass or energy exchange systems) and
nonconservative force systems or dissipative systems. On this basis, we have
given the generalized Lagrange function, it has to do with the kinetic
energy, potential energy and the work of nonconservative forces to do. With
the Feynman path integration, we have given the non-Hermitian quantum theory
of the nonconservative force systems. Otherwise, we have given the
generalized Hamiltonian function for the particle exchanging heat with the
outside world, which is the sum of kinetic energy, potential energy and
thermal energy, and further given the equation of quantum thermodynamics.

\begin{acknowledgments}
This work was supported by the Scientific and Technological Development
Foundation of Jilin Province (no.20190101031JC).
\end{acknowledgments}

\end{document}